\documentclass[5p,times]{elsarticle}

\usepackage[T1]{fontenc}

\usepackage[latin5]{inputenc}
\usepackage{times}
\usepackage[figuresright]{rotating}
\usepackage{float}
\usepackage{subcaption}
\usepackage{url}
\usepackage{setspace}
\usepackage{balance}

\usepackage{multicol}
\usepackage{multirow}
\usepackage{color}
\usepackage{graphicx}
\usepackage[implicit=false]{hyperref}
\usepackage{epsfig}
\usepackage{amsmath, amssymb}
\usepackage{comment}

\usepackage{capt-of}
\usepackage{multirow}
\usepackage{array}
\usepackage{caption} 
\usepackage{pslatex}
\usepackage{enumitem}

\usepackage{tcolorbox}
\usepackage{tabularx}
\tcbset{tab2/.style={colback=red!5!white,colframe=red!50!black,colbacktitle=red!40!white,
coltitle=black,center title}}

\usepackage{colortbl}

\usepackage{amssymb}

\journal{Computer Networks}

\begin{document}

\begin{frontmatter}



\title{Security and Privacy vulnerabilities of 5G/6G and WiFi 6: Survey and Research Directions from a Coexistence Perspective}


\author[inst1]{Keyvan Ramezanpour}

\affiliation[inst1]{organization={Marconi-Rosenblatt AI/ML Innovation Laboratory, \\ANDRO Computational Solutions, LLC},
            city={Rome},
            postcode={13440}, 
            state={NY},
            country={USA}}

\author[inst1]{Jithin Jagannath}
\author[inst1]{Anu Jagannath}

\begin{abstract}
Spectrum scarcity has been a major concern for achieving the desired quality of experience (QoE) in next-generation (5G/6G and beyond) networks supporting a massive volume of mobile and IoT devices with low-latency and seamless connectivity. Hence, spectrum sharing systems have been considered as a major enabler for next-generation wireless networks in meeting QoE demands. Specifically, the 3rd generation partnership project (3GPP) has standardized coexistence of 4G LTE License Assisted Access (LAA) network with WiFi in the unlicensed 5 GHz bands, and the 5G New Radio Unlicensed (NR-U) with WiFi 6/6E in 6 GHz bands. While most current coexistence solutions and standards focus on performance improvement and QoE optimization, the emerging security challenges of such network environments have been ignored in the literature. The security framework of standalone networks (either 5G or WiFi) assumes the ownership of entire network resources from spectrum to core functions. Hence, all accesses to the network shall be authenticated and authorized within the intra-network security system and is deemed illegal otherwise. However, coexistence network environments can lead to unprecedented security vulnerabilities and breaches as the standalone networks shall tolerate unknown and out-of-network accesses, specifically in the medium access. In this paper, for the first time in literature, we review some of the critical and emerging security vulnerabilities in the 5G/WiFi coexistence network environment which have not been observed previously in standalone networks. Specifically, independent medium access control (MAC) protocols and the resulting hidden node issues can result in exploitation such as service blocking, deployment of rogue base-stations, and eavesdropping attacks. We study potential vulnerabilities in the perspective of physical layer authentication, network access security, and cross-layer authentication mechanisms. This study opens a new direction of research in the analysis and design of a security framework that can address the unique challenges of coexistence networks. 

\end{abstract}




\end{frontmatter}


\section{Introduction}
\label{sec:sample1}
The explosion of data generated by a wide range of heterogeneous devices including smartphones, mobile computers, IoT devices, autonomous vehicles, and smart infrastructure, has been the main driver for 5G network development \cite{shen2020internet, jagannath2019machine, yao2019artificial, liu20206g}. This data-centric view of communication networks has resulted in service based architecture (SBA). The SBA allows cloud-based implementation of network functions which facilitates data management while improving scalability and programmability in beyond 5G (B5G) networks \cite{yousaf2017nfv}. The initial architecture of networks has been optimized to achieve high quality of experience (QoE) as the predominant performance metric in the literature of 5G networks \cite{mitola2014accelerating}. However, security architectures have not adapted at the same pace as the new wireless technologies introduced to support QoE demands. This may open the way for serious security breaches either in the form of new security threats or broadening the attack surface for existing vulnerabilities.

Seamless connectivity with low latency and high data rate to a large volume of heterogeneous devices are often considered as the distinctive characteristics of 5G networks. The QoE aims at evaluating the performance of 5G networks in terms of these requirements. A commonly accepted notion about the QoE is the timely delivery of content based on the needs of users. Hence, it is a higher-level objective than the traditional quality of service (QoS) which is characterized by metrics such as data rate and latency of the link provided to a user. While the definition of QoE is broad, without a consensus on a systematic metrics of measure, we note that security and privacy is also important aspect of a user experience. In this view, we can consider the problem of network optimization as maximizing QoE in the sense of content delivery with respect to the desired QoS and with the security and privacy as the constraints of the problem.

The multiple radio access technology (RAT) is a prominent example of distinctive features of 5G networks aiming at high QoE. However, spectrum scarcity for various applications with different RF propagation range requirements is a major challenge. Re-allocation of underutilized spectrum bands is an extremely timely process, faces the resistance of incumbent users, and might interfere with critical military and governmental usage. Spectrum sharing is a promising solution for spectrum scarcity and is considered a major driver for B5G networks in achieving high QoE.

Early examples of spectrum sharing in the U.S. include the commercial use of TV white space (TVWS) spectrum (which is a location-based sharing) and the Citizens Broadband Service (CBS) sharing the 3550-3650 MHz band with incumbent naval radar and fixed satellite systems \cite{mitola2014accelerating}. Spectrum sharing between WiFi and 4G cellular networks in the unlicensed 5 GHz bands has also been standardized by the 3GPP for LTE License Assisted Access (LAA) and enhanced LAA \cite{chen2016coexistence}. The Federal Communications Commission (FCC) of the U.S. and the European Commission have also approved spectrum sharing in 6 GHz unlicensed bands. Hence, 3GPP has defined spectrum sharing specifications, in this so-called greenfield spectrum, for New Radio Unlicensed (NR-U) in 5G networks co-existing with WiFi 6 (based on IEEE 802.11ax specifications) and WiFi 6E (networks operating in 6 GHz bands) \cite{naik2021coexistence}.


Existing network security architectures are designed and developed based on the assumption of independent standalone networks which own the entire network resources, from spectrum to the infrastructure. For clarity, in the paper a standalone network refers to a network infrastructure (base-station and user devices) with exclusive access to the spectrum and without any out-of-network transmissions from coexisting entities. In this security model, any access to the spectrum and resources, communication traffic and network activities are authenticated and authorized within the security framework of a single network. However, the emergence of 5G networks, leveraging software-defined networking (SDN) and network slicing required sharing of network infrastructure among multiple operators and service providers, with different security policies and privacy requirements. Hence, interoperability between various security systems at the level of the core network has become a challenging issue. Similarly, coexistence of networks (WiFi and 5G), and next-generation spectrum sharing systems in general, demands sharing the spectrum among multiple private entities. Therefore, the tolerance of out-of-network activities in the security model at the level of access network is also critical.

Unprotected spectrum sharing in coexistence network environments provides potential adversaries with a covert channel that cannot be detected by existing security mechanisms in standalone networks. The covert channel opens a new surface of security attacks on the networks for which no protection mechanism exists. The security systems of standalone networks observe and respond to intra-network activities while spectrum sharing procedures involve out-of-network spectrum accesses. Hence, spectrum sharing without a security mechanism allows an attacker to exploit the covert channel in deploying new security attacks and/or existing known attacks with higher intensity and simpler implementation mechanisms.

In this paper, we study the security challenges and exploits in the physical layer and access network with an emphasis on the coexistence perspective. To the best of our knowledge, this is the first study in the literature that highlights and focuses on the implications of coexistence on the security of wireless networks. To understand what types of security attacks an unprotected spectrum sharing can facilitate and/or intensify, we first review some of the existing vulnerabilities in standalone 5G and WiFi networks. Next, we will discuss a few security challenges that can emerge because of an unprotected spectrum sharing. Finally, we discuss the challenges of secure coexistence with cryptographic proofs while preserving standalone network privacy.

\section{Vulnerabilities in 5G and WiFi 6/6E}
The security and privacy are intertwined concepts in wireless networks. Privacy refers to inference of information about users by passively observing transmitted signals \cite{clark2020optimizing}. This information can simply include the location and network traffic of users. Passive eavesdropping attacks in the literature are equivalent to privacy attacks on wireless communications, especially at the physical layer. Eavesdropping attacks usually refer to physical layer attacks in wireless communications while privacy is a more generic terminology, mainly used for databases. In this paper, we use these two terminologies interchangeably as the focus is vulnerabilities in wireless communications due to spectrum sharing. 

In the context of wireless communications, security often refers to active attacks, e.g., adversaries introducing elevated interference or intelligent jamming signals for manipulating user transmissions. Prominent examples of active attacks include forcing devices to use alternative data channels, e.g. changing the direction of the beams in MIMO beamforming systems or changing the frequency channel by jamming alternative bands. These attacks can in turn be used to deploy MitM, rogue base-station, DoS, etc. The focus of this survey is spectrum sharing vulnerabilities which mainly rises security/privacy issues at the physical and MAC layers of wireless networks. 

\subsection{Physical Layer Security} \label{sec:physecurity}
Cryptographic proofs for secure communications, in existing standards, are provided by security protocols in different layers of the communication protocol stack. Such security provisions start with authentication (for user/device identification), key agreement protocols, and channel encryption at layer 2 (link layer in OSI model). While the security at the physical layer has been an active research area, standardized frameworks lack security proofs at this layer due to challenges such as variability and uncertainties in the RF propagation channel, device variations, and distributed secret key management for a massive volume of devices before identification (authentication).

A classic method of realizing encrypted physical layers is using spread spectrum systems, either in time or frequency domains. In direct sequence spread spectrum (DSSS), the time-domain samples of a transmitted signal are encoded with spreading codes that have a length much larger than a bit period. If the spreading codes are secret, or encrypted, the DSSS system provides authentication and confidentiality at the physical layer in addition to jammer resilience and anti-spoofing properties. A prominent example of encrypted DSSS physical layer is the Y-code and M-code military signals of the Global Positioning System (GPS)~\cite{barker2000overview}.

Multi-carrier spread spectrum (MCSS) is the equivalent of the DSSS in the frequency domain. The MCSS systems have been popular mainly due to their ability in taking advantage of both orthogonal frequency division multiplexing (OFDM) and code division multiple access (CDMA) in spectral efficiency and robustness to multipath fading and interference \cite{bury2003diversity, li2018index, haab2020performance}. In an MCSS physical layer, the subcarriers of an OFDM signal are encoded with spreading codes. A similar approach can be used to realize an encrypted physical layer by encrypting the samples of the baseband signal in the frequency domain, or subcarriers of OFDM \cite{zhang2016design, sakai2017intrinsic, mirsky2019physical, jacovic2020physical}. 

\subsubsection{Information-Theoretic Security}
Channel coding has become an inevitable component of the physical layer in the communication protocol stack for enhancing and consolidating link reliability. The 3GPP specifications use low density parity check code (LDPC) and polar code for data and control channels, respectively, in the 5G enhanced mobile broadband (eMBB) networks. These coding schemes have also been shown to provide an \textit{information-theoretic security} in a wiretap channel model \cite{wu2018survey, athanasakos2020strong, sasaki2019wiretap, sreekumar2020secrecy}.

Classical ciphers and cryptographic algorithms (both public-key and symmetric) provide security proofs based on the infeasible computational complexity of an eavesdropper who attempts in decoding an encrypted message without a knowledge of the secret key. The information-theoretic security relies on different channel conditions observed by a legitimate receiver and an eavesdropper. The secrecy capacity is defined as the difference between the (information-theoretic) capacity of the communication channels from the transmitter to the intended receiver and from the transmitter to an eavesdropper. Intuitively, if the channel capacity observed by the legitimate receiver is larger, it can carry information which is not received by the eavesdropper regardless of its available computational capability. 

The information security provided by channel coding schemes are evaluated based on either strong or weak secrecy. The strong secrecy is obtained if the mutual information between the transmitted codeword and the received message by the eavesdropper tends to zero (for asymptotically long codewords). The weak secrecy refers to the condition that the average mutual information per bit of the codeword tends to zero. A coding scheme that achieves the eavesdropper channel capacity can also provide perfect secrecy. Based on this relation, designs of LDPC codes have been introduced in \cite{thangaraj2007applications} and \cite{rathi2012performance} that achieve weak secrecy. Further, LDPC codes introduced in \cite{subramanian2011strong} provide strong secrecy when the channel observed by the intended receiver is noiseless. All these schemes assume a binary erasure channel (BEC) model for the eavesdropper channel.

Polar codes have also been shown to achieve weak secrecy in a binary memory-less channel when the main channel of the eavesdropper is not stronger than the main channel of the intended receiver \cite{mahdavifar2011achieving, hof2010secrecy, andersson2010nested}. A multi-block polar coding scheme has been introduced in \cite{csacsouglu2013new} that achieves strong secrecy in addition to reliability in the binary memory-less channel. Further, polar codes in \cite{gulcu2016achieving} and \cite{chou2016polar} achieve the capacity region of a broadcast channel with confidential message (under discrete memory-less model) while providing strong secrecy. The concatenation of polar-polar and polar-LDPC codes are also investigated in \cite{renes2013efficient} and \cite{zhang2014polar}, respectively, for achieving minimum gap to the secrecy capacity.

Since the information-theoretic secrecy relies on communication channel conditions, it can be manipulated by an attacker intruding into the RF environment. Even a passive eavesdropper can gain advantage over a legitimate receiver by using more advanced RF signal processing techniques. Although secrecy capacities are theoretical limits on the information flow in a channel, they still depend on the received signal to interference and noise ratio (SINR). The received signal quality also depends on antenna gain, RF front-end noise figure, signal synchronization, filtering, and decoding algorithms. Hence, an eavesdropping receiver may establish a communication channel with a quality close to the legitimate receiver by employing higher quality RF front-end circuitry, beamforming with higher gain antennas and more advanced, possibly with higher computational complexity, signal processing algorithms. In this perspective, we note that the information-theoretic secrecy is not completely independent of the receiver complexity.

An active eavesdropper can also severely impact channel secrecy by degrading the channel conditions for the legitimate receiver. The eavesdropper can simply increase the interference level at the legitimate receiver while maintaining the channel conditions for its own receiver. Such an attack can be implemented by using self-interference cancellation techniques \cite{hong2014applications, ahmed2015all} or beamforming for targeted interference at the legitimate receiver. We note that increasing interference levels is especially facilitated in coexistence network environments in which higher levels of interference from unknown sources (from co-existing networks) shall be tolerated. Hence, an active eavesdropper can reduce the secrecy capacity to zero.

An alternative active eavesdropper may degrade estimation of the channel state information (CSI). The channel coding schemes require a perfect knowledge of the CSI in providing the promised secrecy rate. In most wireless networks, the CSI is estimated using pilot signals during periodic training phases. The active eavesdropper may interfere only with the training phase, in an attack called pilot contamination, to degrade channel estimation \cite{huang2018pilot, akgun2018vulnerabilities}. The pilot contamination attack in \cite{zhou2012pilot} on time duplex division (TDD) networks has shown to reduce the secrecy rate of the downlink transmissions to near zero. 

\begin{figure*}[t!]
	\centering
		\includegraphics[width=0.8\textwidth]{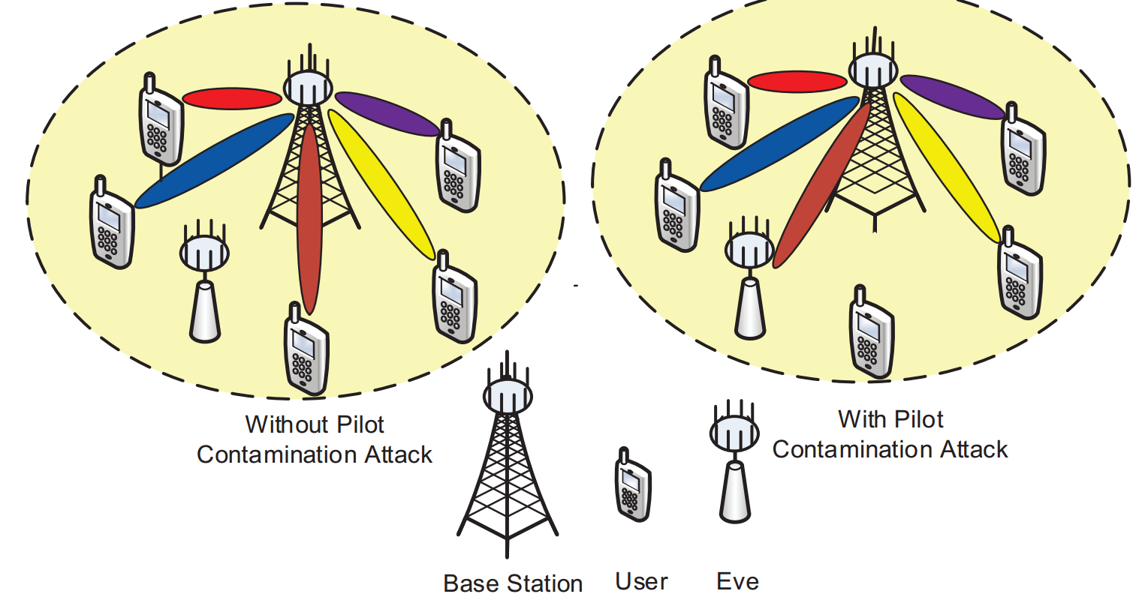}
	\caption{Pilot contamination attack in which an active eavesdropper causes the base-station beam to be directed toward the attacker (Eve) rather than the intended user \cite{wu2018survey}.}
	\label{fig:pilot_contamination}
\end{figure*}

\subsubsection{Beamforming} \label{sec:beamforming}
Beamforming is one of the main characteristics of the new radio (NR) physical layer in 5G networks. The main promise of beamforming is providing significantly higher data rates by enhancing link reliability. Beamforming increases SINR (hence higher reliability) by providing large antenna gains, alleviating multipath fading by forming direct line-of-sight channels, and reducing interference due to space division duplexing. However, a body of research has also made claims on enhancing physical layer security using beamforming. Intuitively, beamforming allows constructing highly directional (thus secret) channels between a transmitter and the intended receiver with minimal leakage of information (signal power) to other directions.

In the view of information-theoretic secrecy, the highly directional transmission of signal power, and reduced multipath fading effects, with beamforming results in consolidated channel conditions for the intended receiver while a significantly degraded channel for an eavesdropper residing at a different direction. Hence, beamforming increases the secrecy capacity and potentially improves data security. As an example, simulation results of \cite{wang2014secure} show a multi-gigabit per second secrecy rate in millimeter wave communications with beamforming. 

An eavesdropper can also exploit beamforming to form a man-in-the-middle (MitM) position by constructing secret channels with legitimate transmitters and receivers. Such an attacker is difficult to detect by legitimate network users. This is in contrast to omni-directional communications where the transmissions from both the eavesdropper and the legitimate transmitter can be detected by the corresponding receiver. In this case, analyzing the transmission patterns can reveal the presence of an active attacker. However, beamforming can potentially facilitate the MitM attack positions.

Apart from MitM attackers, the reliability and secrecy capacity achieved by beamforming can be significantly degraded with pilot contamination attacks in a similar way as CSI estimation. In this case, an active eavesdropper causes the directional beam to deviate from the direction of the intended receiver toward its own receiver using the pilot contamination attack. This attack is depicted in Fig.~\ref{fig:pilot_contamination}. The attacker (Eve) causes interference during the training (channel estimation) phase at the base-station. As a result, the base-station beam is directed toward Eve rather than the legitimate user. In this condition, the advantage of the channel conditions observed by the legitimate receiver over the eavesdropper diminishes and the secrecy rate reduces significantly \cite{wu2016secure, basciftci2015securing}. Further, the higher channel capacity provided to the intended receiver also degrades due to the deviated beam direction.

An alternative passive eavesdropper might exploit the reflections of directional beams to compromise data security provided by beamforming. It has been shown that the reflections of highly directional millimeter waves can be exploited by an eavesdropper to reduce the secrecy capacity significantly \cite{steinmetzer2015eavesdropping}. This work conducts experiments on a millimeter wave software-defined radio (SDR) platform to show that centimeter-scale objects or metal surfaces of devices, such as mobile phones or laptops, can generate reflections with sufficient strengths to reduce the secrecy capacity by 32\%. Further, in the presence of small signal blockage at the direction of the intended receiver, the secrecy capacity can reduce to zero.

\subsection{5G Access Security}

\begin{figure*}[h!]
	\centering
		\includegraphics[width=0.8\textwidth]{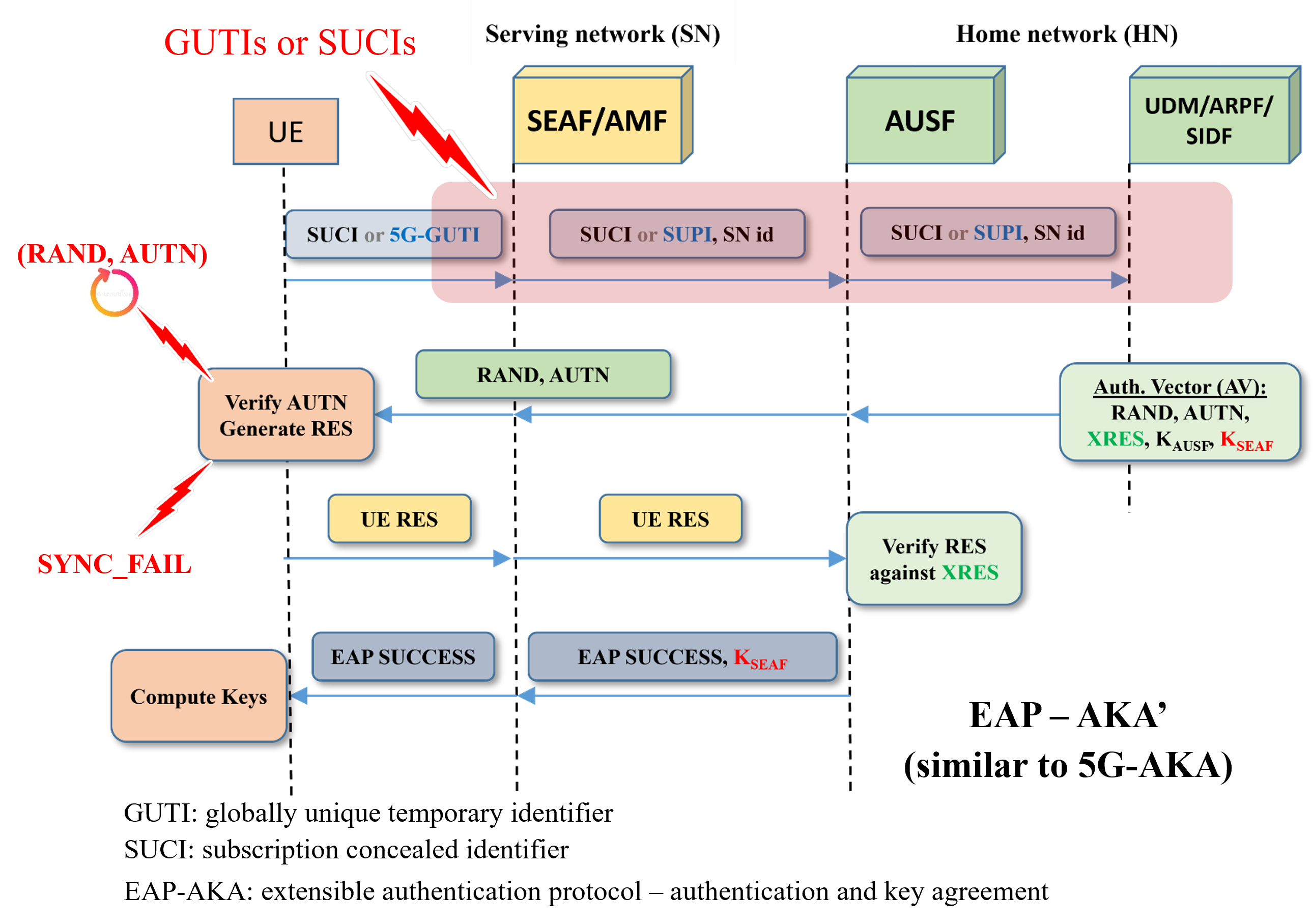}
	\caption{Extensible authentication protocol (EAP) based access security in 5G networks and different attack points on the protocol.}
	\label{fig:eap_aka}
\end{figure*}

The access security in wireless networks is protected by authentication and key agreement (AKA) protocols which constitute the basis of layer 2 security. The 3GPP standards defined an extensible authentication protocol (EAP), called EAP-AKA', for integrating non-3GPP access networks with 4G networks. A widely used non-3GPP access network is WLAN which also employs EAP framework in the WiFi Protected Access (WPA2). The EAP-AKA' is also one of the supported access security mechanisms in 5G networks in addition to 5G-AKA which is a very similar protocol \cite{cao2019survey, ahmad2019security}. The main difference is that SEAF of the serving network also verifies the UE response before sending it to the home network in 5G-AKA. The overall flow of EAP-AKA' is shown in Fig. \ref{fig:eap_aka}. In the following, we review several attacks on the access network security. 

\subsubsection{Identity Confidentiality} 

Different generations of cellular networks have put a large effort in protecting the identity of users. The International Mobile Subscriber Identity (IMSI) is the unique identity of a SIM card. An equivalent identifier in 5G networks is Subscriber Permanent Identifier (SUPI) which is not limited to cellular services and can be used in different 5G environments such as IoT networks \cite{behrad2019survey}. To protect users' privacy, 5G networks employ Subscription Concealed Identifier (SUCI) which contains an encrypted version of SUPI using the public key of a user's home network. After the first attachment to the network (in which the UE transmits SUCI) and initiating the radio channel encryption, the UE is assigned a Globally Unique Temporary Identifier (GUTI) to prevent frequent transmissions of SUCI. 

While defining temporary identifiers (such as GUTI in 4G/5G networks) reduces the chances of exposing SUCI due to frequent transmissions, the values of GUTI have still long lifetimes. Hence, exposing GUTI, e.g., by eavesdropping on the communication channel, can provide an adversary with a soft identity of a user/device. Especially, in a combined attack where the location confidentiality of a user is also compromised in addition to the GUTI, an adversary can attack the privacy of a user and get access to such information as phone number, network activity, calls, and SMS. Further, the attacker can track the user even if the GUTI value is refreshed. Hence, disclosure of GUTIs with long lifetime can have a similar effect as IMSI catching attacks.

Although 5G networks employ SUCI and GUTI to protect the permanent identity of devices, it is still possible for a UE to transmit SUPI (or equivalently IMSI) in plaintext in 5G networks. In case of unauthenticated emergency calls, the security of SUPI is not guaranteed and the UE may transmit plaintext SUPI. The emergency services are available to UEs that fail authentication and scenarios where authentication cannot or may not be performed. In a coexistence network environment, an attacker can actively access spectrum, and introduce faults, without being identified as adversary. Hence, by generating an emergency scenario, e.g., causing authentication failure by introducing faults, an attacker can force a UE to transmit SUPI/IMSI in plaintext. By eavesdropping on the communication channel, the attacker can obtain the IMSI. Alternatively, the attacker can set up a rogue base-station, masquerading a legitimate gNB, and make the UE to transmit IMSI directly to the attacker in an emergency service request.

\subsubsection{International Mobile Subscriber Identity (IMSI) Cracking}

Although the identity used in 5G networks (SUCI) is encrypted, an adversary can still crack the concealed IMSI using a combination of different techniques. The IMSI is a 49-bit identifier in which 18 bits are common country codes known to potential attackers. Further, a side-channel attack called ToRPEDO (TRacking via Paging mEssage DistributiOn) can be used to recover 7 bits of the IMSI with less than 10 calls even under the assumption that the Temporary Mobile Subscriber Identity (TMSI) changes after every call \cite{hussain2019privacy}. The attack works based on tracking the paging occasions of a device (periodic polling of a device for pending services in the low-power idle state). The time period of the paging occasions is fixed for a cellular device related to 7 bits of the IMSI. Hence, an attacker can verify whether a device is in the vicinity (a coarse-grained location information) by observing the timing of the paging occasions along with the corresponding 7 bits of the IMSI.

The remaining 24 bits of the IMSI can be obtained using a brute-force attack exploiting security weaknesses of the 5G authentication mechanism. Having the public key of the home network, an attacker can forge a SUCI by encrypting a guess for the IMSI and sending it to the core network for identification. The response of the network is either AUTH\_REQUEST (identity is valid in the network) or REGISTRATION\_REJECT (invalid identity). If the AUTH\_REQUEST response is received, then it is forwarded to the device to verify whether the guessed SUCI belongs to the victim device. The response of the device is either AUTH\_RESPONSE (identity belongs to the device) or AUTH\_FAIL (incorrect identity). A real attack leveraging ToRPEDO has shown to be successful in recovering the IMSI in 74 hours.

\subsubsection{Location Confidentiality}
\label{sec:traceability}
The ToRPEDO attack discussed above can provide a coarse-grained location information about a cellular device. The attacker can further employ RF signal processing techniques, such as angle-of-arrival (AoA) estimation and receive signal strength (RSS), to obtain and track a fine-grained location of a device. A complementary attack, called traceability attack, can also be used to verify the presence of a specific device (already characterized, e.g., using ToRPEDO) in the vicinity of the attacker. 

The traceability attack exploits a vulnerability in the EAP authentication mechanism of 5G networks as shown in Fig. \ref{fig:eap_aka}. By eavesdropping on the communication of a device during the initial authentication, an attacker binds the challenge message (RAND, AUTN) to the device. To verify the presence and track the location of the device, the attacker can replay the challenge message to the device. If the device is present in the vicinity, it would respond with SYNC\_FAIL message \cite{basin2018formal}. Hence, the attacker can track the user without requiring message exchanges with the core network.

\subsubsection{Denial-of-Service (DoS)}

As 5G networks are expected to provide connectivity to a massive volume of devices, from mobile to IoT devices, DoS and distributed DoS (DDoS) have also become more serious and effective security attacks with easier implementation mechanisms. These attacks fall into the category of unintrusive precision cyber weapons (UPCW) which has emerged as a serious cybersecurity threat in the era of IoT. These attacks often require low pre-attack intelligence gathering and pre-positioning of exploits while inflicting more effective damage on the network performance. The UPCW attacks can exhaust and overload the resources both at the network core (such as authentication servers) and devices as in DDoS, Telephony Denial of Service (TDoS), and Denial of Sleep (DoSL).

Examples of DoS targets on the authentication protocol of 5G networks are shown in Fig. \ref{fig:eap_aka}. An attacker (rogue base-station) can send many authentication request messages to a device and overload the computational resources of the device. On the other hand, if the attacker sends such messages to many (IoT) devices (or force devices by installed malware), then the devices will send their GUTIs to the network core (SEAF in the serving network). In any case, the serving network shall send the SUPI (corresponding to the GUTI) or SUCI to the home network for generating the corresponding authentication vector. Only after verifying the response of the device (RES) against the expected response (XRES) the devices are authenticated. 

If a massive number of authentication requests are transmitted to the network core in a short period, the communication and computational resources of the network will be exhausted. Similarly, if a massive number of authentication requests are sent to a device, it is forced to calculate the response frequently which overloads its computational and power resources. In the case of IoT devices, this causes the depletion of battery, the so-called DoSL attack.

In addition to manipulating devices to overload the network, an adversary can collect a massive number of GUTIs and/or SUCIs and flood the network core with authentication requests. Further, the attacker may send fake SUCIs using rogue or infected devices with malware. As shown in the diagram of Fig. \ref{fig:eap_aka}, the serving network shall send the SUCIs to the home network to decrypt the concealed identifiers and verify the authenticity of the SUCIs. Hence, the resources of network core will be exhausted. In this case, the attacker does not even need communication with devices.

\subsubsection{Handover Security}

\begin{figure*}[t!]
	\centering
		\includegraphics[width=0.8\textwidth]{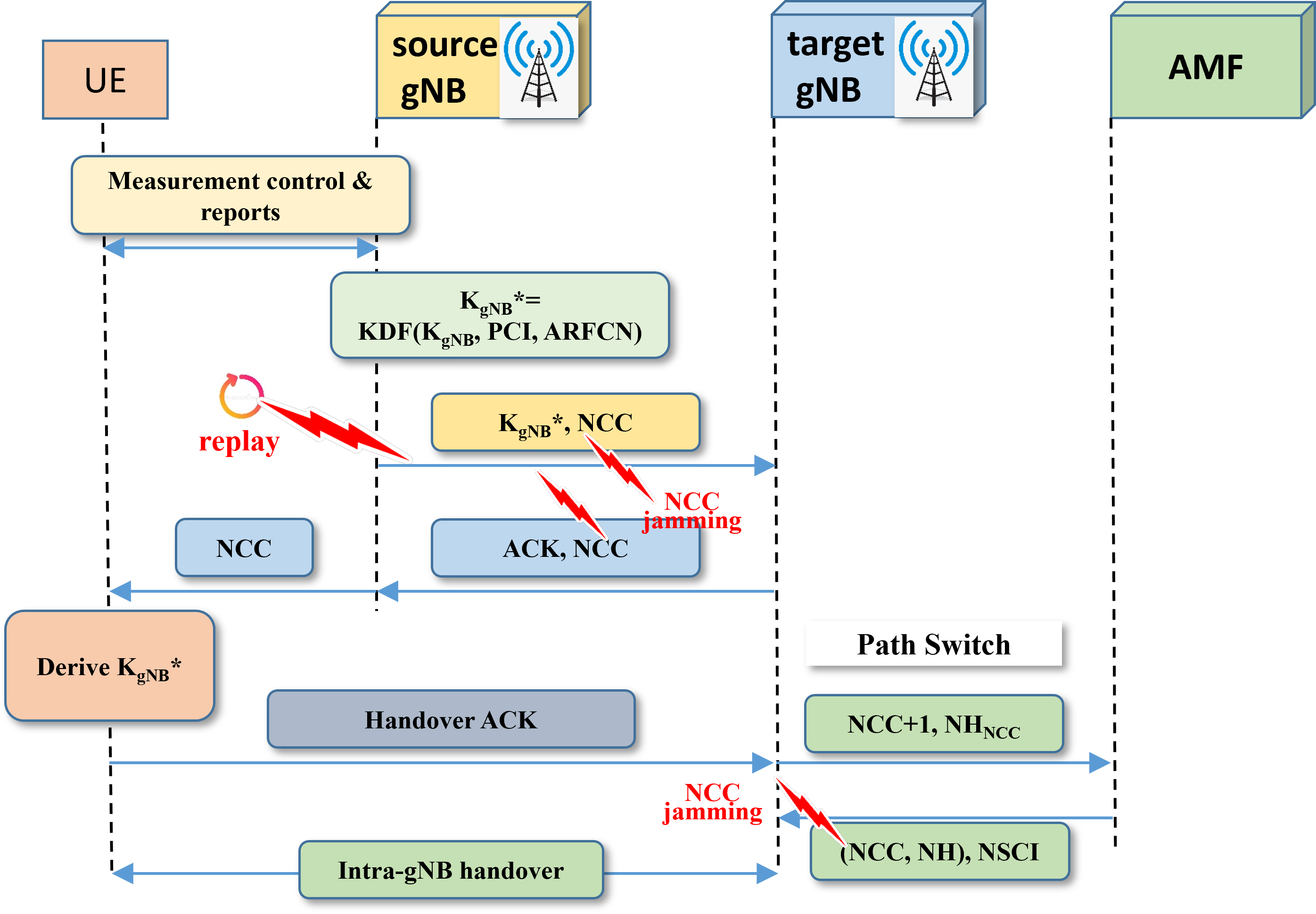}
	\caption{Security protocol for handover in 5G networks with different vulnerable points for replay and jamming attacks.}
	\label{fig:handover}
\end{figure*}

The security of the handover process in dynamic environments is a major challenge for 5G (and beyond) and WiFi (especially wide area and enterprise) networks. Especially in high-mobility applications, the latency of complex authentication and handshake protocols cannot be tolerated. On the other hand, devices are most vulnerable to security attacks, such as rogue base-station and DoS, during handover since they have the weakest connectivity. Existing security architectures exploit the initial authentication process as a trust basis for simplifying the security mechanism of the handover process. While this approach contrasts with the perspective of future zero trust architectures, it fulfills the latency requirements which is critical for delay-sensitive applications.

The security handshake during handover in 5G networks is shown in Fig. \ref{fig:handover}. When a decision is made on handover from the source to the target gNB, based on path detection, channel conditions, and user location, the security handshake is initiated. The source gNB derives the session key $K_{gNB}^*$ from the current key $K_{gNB}$ using a key derivation function (KDF) with physical cell ID (PCI) and absolute radio frequency channel number (ARFCN). The source gNB then sends the new key and the next hop chaining counter (NCC) to the target gNB. This process provides forward security; having the current key does not reveal information about previous session keys. However, backward security is not guaranteed; if an adversary compromises the source gNB, then all future session keys are revealed. To solve this issue, the handover mechanism also includes an intra-gNB process (between user device and target gNB) after the device is switched to the target gNB. However, the intra-gNB process still incurs large communication overhead and computational complexity.

A major vulnerability of the above handover process is the failure due to replay message by a rogue base-station (gNB). The attacker intercepts the first message between gNBs, i.e., $(K_{gNB}^*, NCC)$ and replays this message whenever the UE is going to handover between two gNBs. The target gNB has no means to verify the authenticity of this message. Hence, it uses the received session key with the UE. It also transmits the received NCC back to the UE. However, this NCC is different from the local counter at UE (since it was a replay message) and the handover fails. In a similar attack, called jamming or de-synchronization attack, the adversary can change the value of NCC which again leads to handover failure \cite{cao2013survey}. These attacks are facilitated through a rogue gNB activated during handover.

\subsection{WiFi 6/6E Access Security}

Access security in WiFi networks is also based on the layer 2 security (authentication) like 5G networks. In the third generation, WiFi systems leverage the EAP-based authentication framework in WPA2 while the fourth generation of WiFi employs WPA3 based on Simultaneous Authentication of Equals (SAE) standardized in IEEE 802.11. While WPA2 framework suffers from similar EAP vulnerabilities as in 5G networks, the WPA3 is also susceptible to downgrade attacks, DoS, and side-channel attacks due to high computational complexity. In the following, we briefly review several attacks on the access security of WiFi networks using WPA2 and/or WPA3 protection.

\subsubsection{Rogue Access Point (AP)}

A well-known and effective on the access security of WiFi networks is the rogue AP, commonly referred to as evil twin \cite{abo2018enterprise, shrivastava2020evilscout}. Since the beacon packets of AP are not encrypted, an attacker can easily access the network name (SSID) and its MAC address (BSSID). Hence, the attacker may impersonate a legitimate AP (LAP) and force devices to connect to the rogue AP (e.g., by transmitting with higher signal strength). If a device is already attached to the LAP, the attacker can deploy a de-authentication attack and encourage the device to connect to the rogue AP. 

While the security mechanism of WiFi (especially third generation and earlier) are vulnerable to evil twin attacks, there are also effective protection techniques to detect rogue APs. An example is verifying the duplicate association of a WiFi client (user device) with different APs that happens when both legitimate and rogue APs are connected to the client at the same channel. Further, the security policy rules of the network might prevent a device to communicate with an AP without mutual authentication of the network. However, we note that all these provisions are implementation-specific and do not provide cryptographic guarantees in protecting against different types of attacks.

\begin{figure}[t!]
	\centering
		\includegraphics[width=0.49\textwidth]{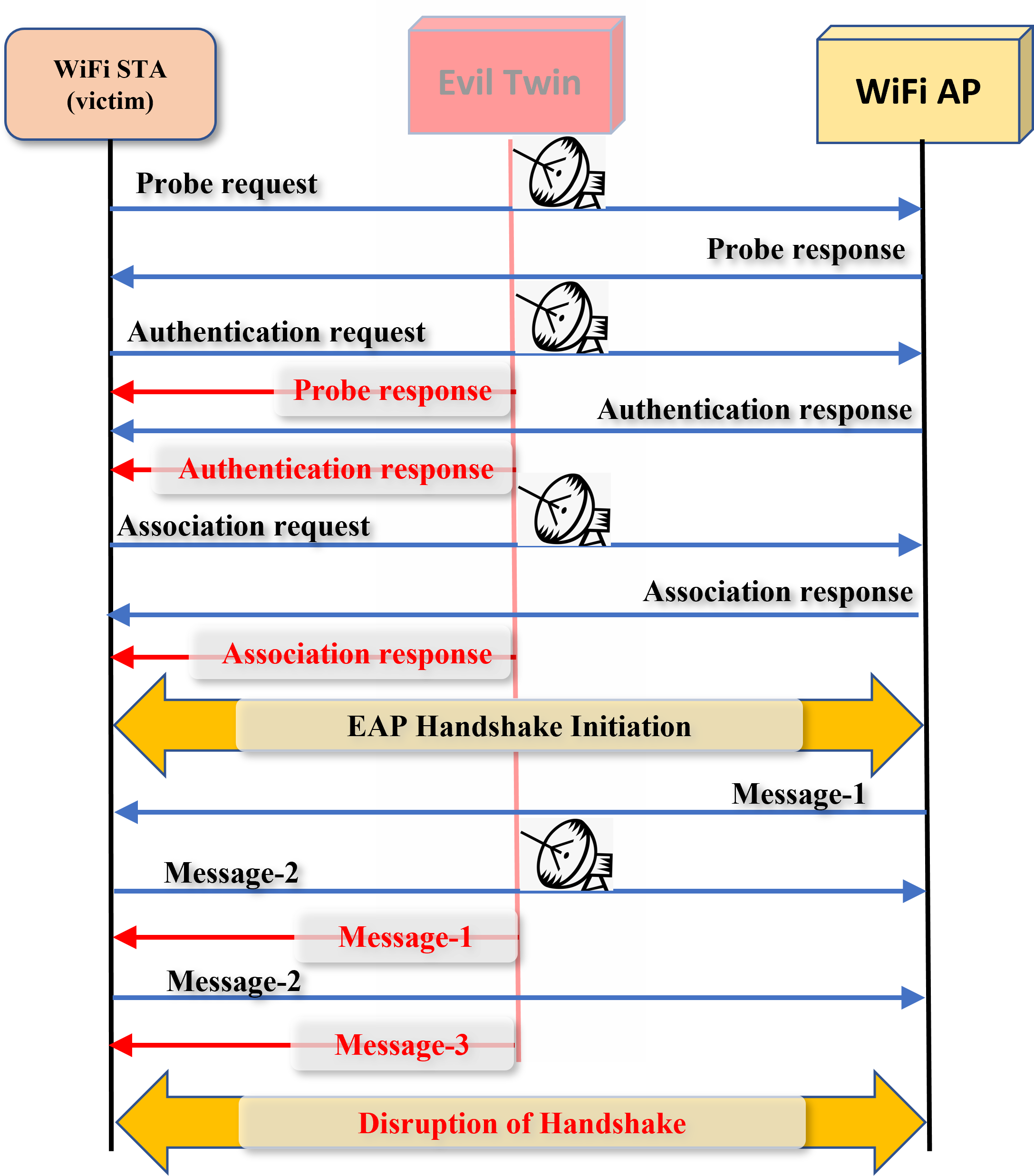}
	\caption{Evil twin assisted service blocking attack attack on EAP-based authentication handshake in WPA2 protected WiFi.}
	\label{fig:wifi_blocking}
\end{figure}

In addition to the man-int-the-middle (MitM) attack, an evil twin can deploy an easier and effective attack on the WiFi access security called service blocking. In this attack, which can also be considered as a DoS attack, the evil twin does not associate with WiFi devices but disrupts the authentication process of WPA2 which results in blocking any connection to the LAP. The diagram in Fig. \ref{fig:wifi_blocking} shows the flow of service blocking attack on the WPA2 authentication protocol of WiFi. The WiFi STA (device) sends probe and authentication requests to the LAP for initializing a connection. The LAP responds with respective responses after every request. As shown in the figure, after sending the authentication request by the client, both LAP and evil twin send the probe response to the client. Similarly, authentication/association response messages are also sent to the client by the LAP and the evil twin at the appropriate timeline as shown in Fig. \ref{fig:wifi_blocking}. The responses of evil twin is shown with red horizontal lines from the mid section of the figure to the WiFi STA. After receiving the authentication response from, the STA initiates the 4-way EAP handshake protocol. Regardless of whether the client receives the message-1 of the EAP handshake first from the LAP or evil twin, it responds with message-2. However, upon receiving another message-1 from the other AP (evil twin or legitimate), the handshake protocol fails and the connection to the LAP is disrupted.

\subsubsection{Key Reinstallation Attacks}

For mutual authentication in WPA security systems, the pre-shared Pairwise Master Key (PMK) is used to generate the session keys called Pairwise Transient Key (PTK) using random numbers SNonce (at AP) and ANonce (at the client). The key reinstallation attack is deployed by replaying message-3 in the handshake protocol of EAP which results in resetting of the nonce and replay counters \cite{vanhoef2017key}. As a result, the previous PTK, already in use, will be installed for subsequent communication. However, for a successful attack, the attacker requires a MitM position in which it blocks message-4 from arriving at the AP before re-transmitting message-3. Depending on the security protocols used, the key reinstallation allows further replay attacks, decryption, and forgery of messages. 

\begin{figure*}[t!]
	\centering
		\includegraphics[width=0.6\textwidth]{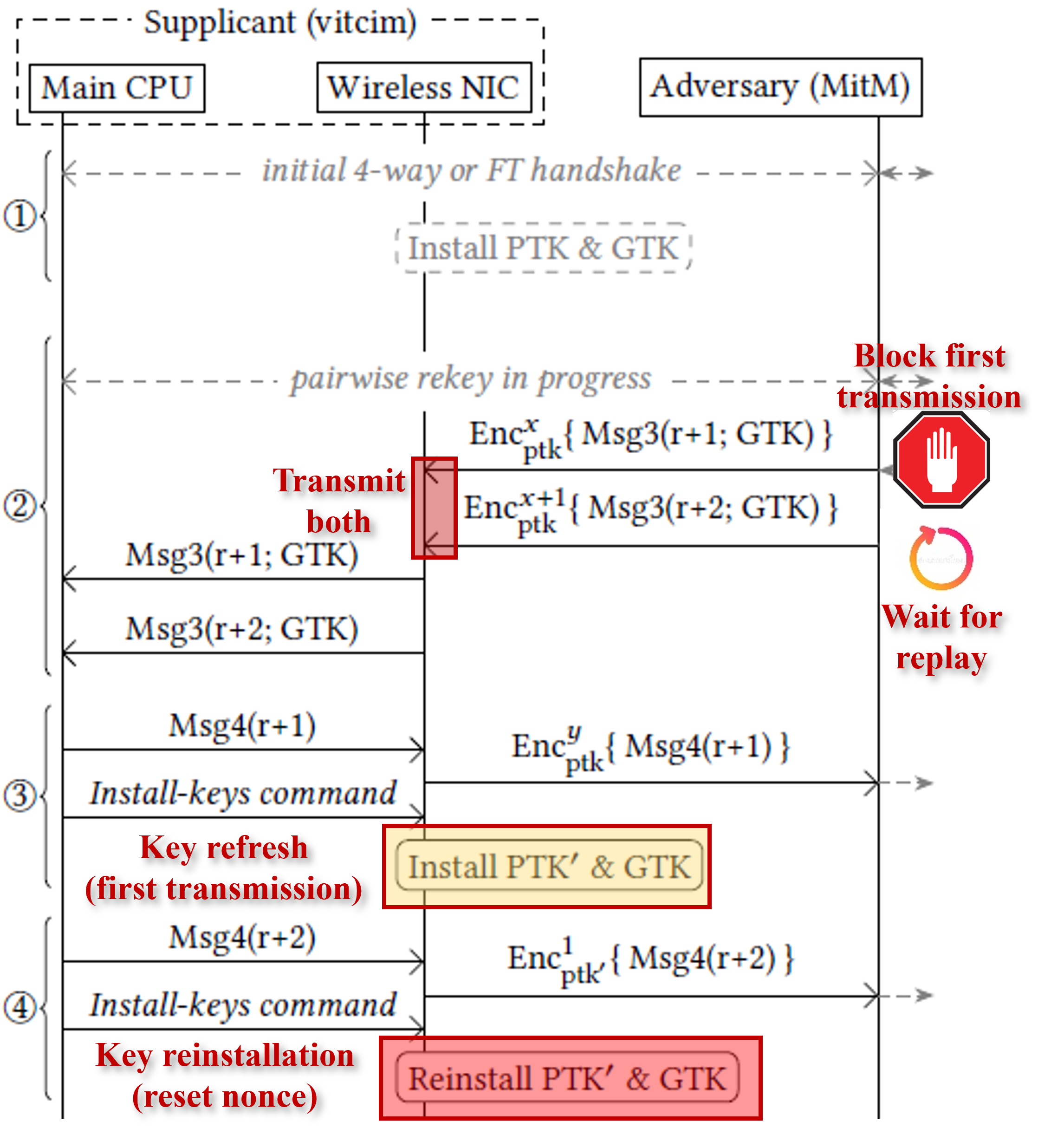}
	\caption{Key reinstallation attack on EAP-based authentication handshake in WPA2 protected WiFi with man-in-the-middle (MitM).}
	\label{fig:wifi_kreinstallation}
\end{figure*}

Despite several implementation-specific provisions in detecting and preventing the replay of message-3, many hardware platforms are still vulnerable to key reinstallation attacks. As an example, most secure implementations might accept only an encrypted version of message-3 in retransmission. Examples include OpenBSD, OS X, and macOS which mandate encryption of message-3. However, a race condition between the components implementing the handshake protocol (e.g., CPU) and data confidentiality protocols (e.g., network interface controller) can still be exploited in a key reinstallation attack during the key refresh operation. 

The flow of key reinstallation attack during key refresh is shown in Fig. \ref{fig:wifi_kreinstallation} in which message-3 is also encrypted. The key refresh exchange happens in a similar way as the 4-way EAP handshake with the difference that the messages are now encrypted with the current key. As shown in the figure, during stage 1, the initial key is established in stage 1 of the diagram. At stage 2 (when a key refresh is required), the EAP handshake with encrypted messages is started. The target of the attack is the (encrypted) message-3 of the handshake. The attacker needs to form a MitM position. In this case, when message-3 (encrypted with the current PTK) is transmitted from the AP to the client, the attacker blocks the message. Hence, the AP retransmits message-3. At this point, the attacker transmits both messages to the client device at once. The wireless network access controller (NIC) decrypts the messages (using the current PTK) and sends them to the CPU. After receiving the first message, the CPU refreshes the PTK. Similarly, the CPU receives the second message (which was encrypted but with the old PTK) and installs PTK again. This causes the nonce value associated with the PTK to restart from 1.

The above key reinstallation attack, even on secure implementations that mandates encryption of message-3 during rekeying, results from the race conditions between different security modules in the system. Specifically, modern NICs support advanced encryption protocols for data confidentiality. However, the separation between different security components enables new security vulnerabilities even though provisions such as mandatory encryption messages are in place.

\subsubsection{DoS Attacks}

In response to key reinstallation attacks, the WiFi Alliance introduced WPA3 which included a variant of Dragonfly handshake protocols, based on the SAE framework, in the WiFi security system \cite{vanhoef2020dragonblood}. Further, it defines a transition mode in which both WPA3 and WPA2 are supported for backward compatibility. While the SAE-based WiFi handshake protocol promises improved security, it incurs large computational overheads leading to DoS attacks. Hence, the implementation of WPA3 security on the commercial off-the-shelf (COTS) is challenging.

The Dragonfly handshake of WPA3 supports both elliptic curve cryptography (ECC) and finite field cryptography (FFC) for key derivation from a pre-shared key/password and mutual authentication. The Dragonfly protocol uses a try-and-increment loop mechanism to convert the Hash of the password to a valid point on the elliptic curve (or the multiplicative group). To prevent timing attacks, a large number of operations are required in the process (order of magnitude larger than alternative methods). Hence, Dragonfly also employs an anti-clogging mechanism to prevent attackers from deploying DoS attacks by exploiting the large overhead. However, the anti-clogging mechanism does not guarantee protection against DoS attacks. In an experiment, a Raspberry Pi B+ with a 700 MHz CPU was used as an adversary station attacking a professional AP with a 1200 MHz CPU. The results of the experiment showed that an attacker can increase the CPU usage of the AP to 100\% by spoofing only 8 commit exchanges of the Dragonfly protocol per second.

\subsubsection{Downgrade Attacks}

In the transition mode, both WPA3 and WPA2 are supported in which the respective authentication protocols of the WPA3/WPA2 use the same password. Hence, by deploying a downgrade attack (e.g., forging beacon messages and forcing WiFi stations to use WPA2), the password can be recovered by attacking the WPA2 security protocols. To prevent the downgrade attack, the WPA2 handshake in the transition mode incorporates a Robust Security Network Element (RSNE) with a list of all supported protocol suites. Hence, the client device can detect a forged beacon message from an adversary. 

The above defense mechanism is still vulnerable to downgrade attack. The attacker can transmit a beacon with WPA2-only network with the SSID of the legitimate AP (with WPA3 support) to the client. Since the first message of the handshake is not authenticated, the client connects to the attacker's AP and sends the second message which is authenticated. At this point, the attacker can use the second message in an offline dictionary attack to recover the password. In this attack, the adversary does not even need a MitM position. 

Another downgrade attack targets the set of elliptic curves or multiplicative groups. The SAE framework defines different groups which are prioritized and configured by the user. The negotiation mechanism for choosing a group can be exploited by an attacker to force a particular group more favorable to a specific attack. This attack is deployed in a MitM position in which the attacker can block some negotiation exchanges and only allow those messages corresponding to a preferred group. This is especially important considering that different groups might exhibit different types of vulnerabilities against side-channel attacks such as timing and cache attacks. Hence, the attacker can force users to choose a group that is most vulnerable to such attacks.

\section{How  Coexistence and Unlicensed Spectrum Sharing Exacerbates Security Challenges} 
\label{sec:coexistence}
\begin{figure*}[t!]
	\centering
		\includegraphics[width=0.8\textwidth]{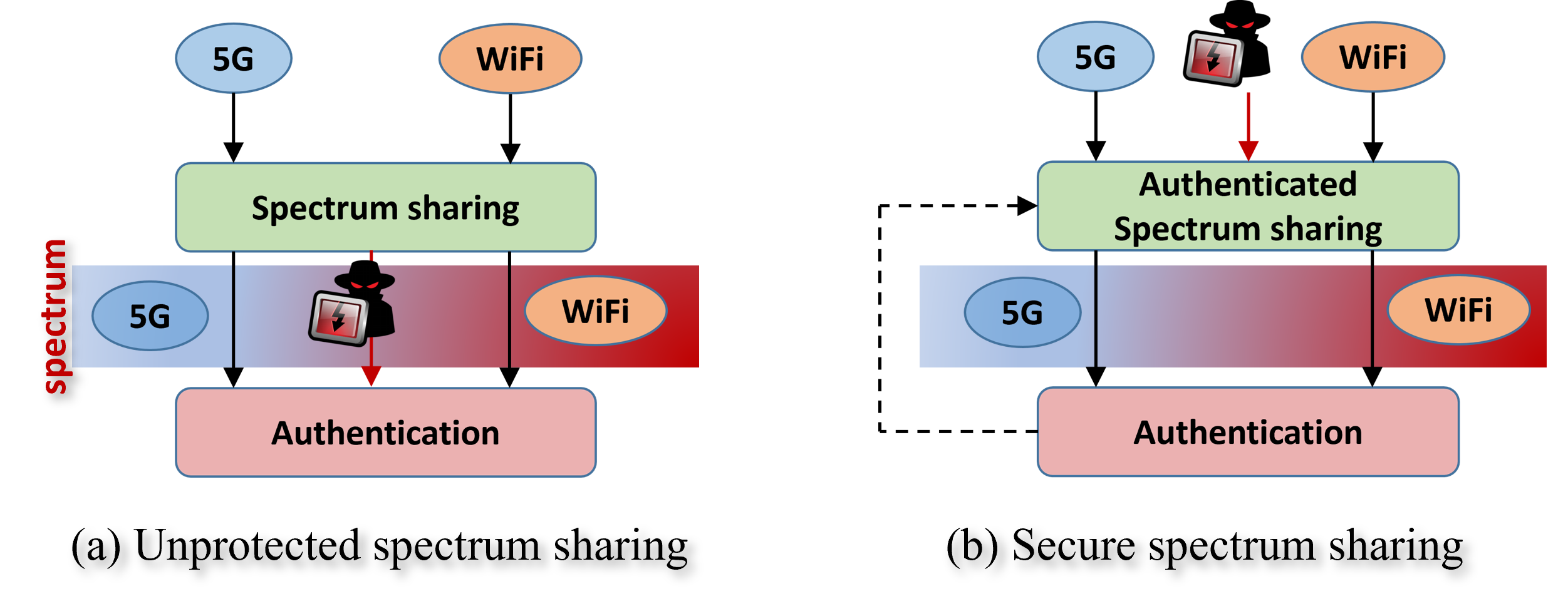}
	\caption{Comparison of (a) unprotected and (b) secure spectrum sharing in allowing an attacker to access (hijack) the spectrum.}
	\label{fig:sharing}
\end{figure*}

Spectrum sharing in coexistence network environments adds an additional surface of attack to the security system of standalone networks. The first layer of security in standalone networks is the authentication framework (layer 2) which assumes any spectrum access belongs to the same network and must be authenticated. The major challenge of spectrum sharing originates from the fact that existing security frameworks do not recognize out-of-network accesses. However, in a coexistence environment, the network entities (base-stations or devices) shall first compete for the spectrum with other networks (and most likely with independent and private security mechanisms). Only after gaining access, the security framework of a network can authenticate and authorize the access. In this environment, an adversary can compete for the spectrum as a legitimate entity while there is no security mechanism in standalone networks that can detect such an attacker. This vulnerability is conceptually depicted in Fig. \ref{fig:sharing}.

Existing spectrum sharing solutions for coexistence network does not include any security mechanism and focus on the network performance metrics, specifically the quality of experience (QoE). They rather postpone security provisions to the access security frameworks at the upper layers of the communication protocol stack. This causes an immediate vulnerability as shown in Fig. \ref{fig:sharing} (a); coexistence of an attacker with legitimate network users is inevitable in such environments. This coexistence of an attacker is a new attack surface on spectrum sharing systems which does not exist in standalone networks. This view reveals the necessity for a secure mechanism in spectrum sharing as intra-network security mechanisms are only relevant after accessing the spectrum. To emphasize the importance of secure spectrum sharing, we briefly review the opportunities for an attacker to exploit unprotected spectrum sharing and deploy new attacks or implement existing known attacks with larger impact and easier mechanisms.

\subsection{Security in Spectrum Sharing Systems}
Since coexistence of 5G and WiFi networks is an emerging trend, to address the spectrum scarcity in next generation wireless networks, limited research has studied the implication of coexistence for the security of networks. Traditional spectrum sharing (SS) systems are a closely related, and a more traditional, network solution for improving the efficiency of the spectrum usage as opposed to standalone networks (with licensed spectrum bands).

The main difference between traditional SS systems and coexistence network environments is the asymmetric versus symmetric spectrum access in the two schemes, respectively. Traditional SS systems consist of incumbent or primary users (PU) who share the spectrum with secondary users (SU). This is known as a two-tier spectrum sharing mechanism. The priority of spectrum access is always with the PU. The secondary users are allowed to use the spectrum only if no PU is present. While this scheme allows reuse of spectrum in geographical areas or time slots without PU users, it still does not provide a fine-grained spectrum sharing among users of two (or more) networks which might achieve the overall network capacity. 

A coexistence networking scheme provides a fine-grained access to the spectrum, both in frequency and time domains, to the users of two or more networks in a fair (and symmetric mechanism) as described in the next section. Since both networks have symmetric access to the spectrum, the respective users experience less frequent outages which in turn improves the overall network capacity. An study of achievable network throughput in 5G and WiFi coexistence environments is provided in \cite{naik2021coexistence}. However, security implications of coexistence is similar to the traditional SS systems as both schemes share the same frequency bands. Hence, it is expected that similar security challenges as SS systems are also transferred to the coexistence networking schemes. In this section, we review the known security challenges of such SS systems.

A well-known attack on a two-tier spectrum sharing is the primary user emulation (PUE) in which an adversary emulates and transmits the signals of a PU. While this attack can be detected by PUs, using the security mechanisms of the primary network, the SUs do not have mechanisms to verify the legitimacy of the accesses by the PUE. Hence, a PUE attacker can prevent SUs from accessing the spectrum. This is also known as dynamic spectrum access (DSA) DoS attack. The PUE attack can also be exploited to deploy more complicated attacks such as \textit{spectral honeypot} \cite{newman2010case}. In this attack, the adversary forces a SU to use a target channel by occupying other channels. This attack can be used to facilitate man-in-the-middle attack or simply to manipulate the SU to generate more interference in the target channel.

In addition to PUE, the SUs can also generate increased (potentially unwanted) interference for the PUs. If a SU fails to sense the signals of PUs, e.g., due to multipath fading channels, it will access the spectrum which in turn causes harmful interference for the PUs. One solution is using a distributed sensing and centralized decision mechanism in accessing the spectrum by SUs. In this case, all SUs report the results of their spectrum sensing to a centralized spectrum access system (SAS) which authorizes the SUs to transmit (if no PU signal is detected). However, this mechanism is vulnerable to spectrum sensing data falsification (SSDF) attacks in which adversaries impersonate SUs and send false data to the SAS \cite{rawat2010collaborative}. A more comprehensive review of different attacks on two-tier SS systems, based on both sensing and database sharing mechanisms, is available at \cite{park2014security}.

A natural result of spectrum sharing is the potential of eavesdropping attacks (active or passive). To prevent interference in the shared spectrum, the signals of a network must be detectable by foreign users. In this way, the network activity and traffic can also be analyzed by foreign users which violates the privacy of users. A study of privacy exploits, in terms of location information leakage, is provided in \cite{liu2021bilateral}. This work proposes a game-theoretic solution for protecting location information of PU and SU from each other. In addition to location information, adversaries can intercept the signals of legitimate users to extract private and/or sensitive information. One solution in MIMO systems is establishing a data channel with high secrecy capacity using beamforming as described in Section \ref{sec:beamforming}. Intelligent (friendly) jamming signals has also been used in these systems to increase the secrecy capacity. An example is the artificial noise transmitted in the null space of the channel between legitimate communicating pairs \cite{wang2015secure, jiang2022intelligent}. However, these techniques requires an estimate of the channel state information (CSI) which can be the target of an active eavesdropping attack.

Most existing research on the security of spectrum sharing has studied vulnerabilities in multi-tier systems (asymmetric spectrum accesses). Although similar vulnerabilties are also present in coexistence environments, there are specific security issues in coexisting networks with symmetric spectrum accesses. In the following sections, we review part of potential vulnerabilities specific to coexistence environments which also reveals the similarities with multi-tier systems.

\subsection{Spectrum Hijack} \label{sec:hijack}

Since spectrum sharing is happening before any security mechanism in standalone networks is activated, an attacker can masquerade as a legitimate network entity that shares the spectrum without being identified as an adversary. This leads to a spectrum hijack attack that does not have any analogous condition in standalone networks. This coexistence vulnerability severely degrades QoE while the main promise of coexistence networks is providing high QoE guarantees. In addition to QoE degradation, spectrum hijack attacks can be a serious threat to the public safety and mission-critical communications. Seamless connectivity and low latency promise in 5G and WiFi 6/6E networks are the fundamental requirements of technologies dealing with public safety, including autonomous vehicles, smart cities and infrastructure, emergency responders, and surveillance systems. Many of these applications are also delay-sensitive in the sense that on-time delivery of content is the critical requirement for their operation. However, the simplest spectrum hijack attack can disrupt the network connectivity or at least introduce large latency in providing required services.

Spectrum hijack attack is facilitated by independent security mechanisms and privacy requirements in standalone networks. One solution to protect against this attack is using a trusted third-party to provide access security services (i.e., authentication services) for different networks (5G and WiFi 6/6E). In this way, every network can verify the legitimacy of accesses by using access tokens received from the trusted party. However, this approach requires substantial changes in the security architecture of networks which seems impractical. Further, the privacy requirements in standalone networks, in protecting identifiers and network traffic, prevents sharing information of users between networks.

\subsection{Service Degradation}

\begin{figure*}[t!]
	\centering
		\includegraphics[width=\textwidth]{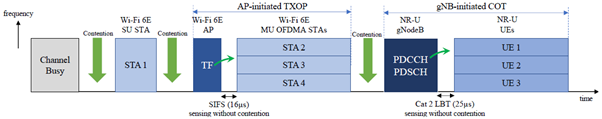}
	\caption{Spectrum access based on listen before transmit (LBT) protocols for coexistence of 5G and WiFi 6/6E \cite{naik2021coexistence}.}
	\label{fig:contention}
\end{figure*}

A coexistence attacker can exploit unprotected spectrum sharing to cause interference between transmissions of different networks which results in quality of service (QoS) degradation (high latency and low throughput). To understand this attack, we consider the primary idea of existing solutions for spectrum sharing in 5G and WiFi 6/6E coexistence as shown in Fig. \ref{fig:contention}. Every wireless device first contends for a free spectrum using a family of LBT-based protocols. The WiFi 6 system supports two modes of single user (SU) and multi-user (MU). In the SU mode, every WiFi station (STA) contends for the spectrum individually. The MU mode of WiFi 6/6E is very similar to the uplink OFDMA of 5G networks. In this mode, the WiFi AP or 5G gNB contends for the spectrum. After the successful acquisition of free spectrum, the AP/gNB schedule their users in OFDMA units.

A major challenge of the spectrum sharing mechanism in coexistence networks is the hidden node issue. If the base-station of one network is a hidden node for the other, most likely both networks schedule their respective users at the same OFDMA units. Hence, the transmissions in the two networks collide with each other. Hidden nodes are detected and/or avoided in standalone networks using medium access control (MAC) protocols. However, in coexistence environments, the inter-network hidden nodes are inevitable simply due to independent MAC protocols of the networks and the absence of information sharing between them. This can arise security issues in a coexistence environment.

The hidden node issue in coexistence networks can be exploited by attackers to severely degrade QoS. A simple attack exploiting the potential existence of hidden nodes can cause a service degradation even if the base-stations are not really in the hidden node position. Assume the base-stations contend for the spectrum and AP/gNB successfully acquire certain channels (e.g., OFDMA units). An attacker may forge the packets of control channel for gNB/AP and schedule the respective users of the network in the same channels that AP/gNB had already acquired. As a result, the users of both networks transmit in the same OFDMA units. Although scheduled users still employ LBT protocols for transmission, this condition can substantially degrade QoS. First, users must wait for the channel to get freed which increases network latency. Second, collision of packets from different networks is very likely especially in wider areas where propagation delay is long. Third, the wireless devices from different networks can be in hidden node positions which results in collisions of packets from different networks.

\subsection{Pseudo Man-in-the-Middle} \label{sec:mitm}

Most of the existing attacks on the access security of wireless networks, as discussed in the previous sections, are either enabled or facilitated by a MitM attacker. A prominent example is the key reinstallation and downgrade attacks on the WPA2 and WPA3 security of WiFi. In these attacks, an adversary requires to selectively block and transmit or replay messages in the authentication handshake protocols. The downgrade attack from WPA3 to WPA2 might not need a MitM position but is more effective by such an attacker with a lower chance of detection. Similarly, the service blocking attack on WiFi using an evil twin can be more effective in a MitM attack position. We discussed that the presence of evil twin results in duplicate association of WiFi devices which can alarm the presence of rogue APs. However, an evil twin in a MitM position is harder to detect. 

The MitM attacker can also cause serious threats to the access security of 5G networks. Examples include replay messages in transmitting a massive number of re-authentication requests and deploying a DoS attack on the UEs or core network. While the MitM attack is not necessary in this case, it can reduce the chance of detecting an attacker by analyzing the spectrum activities at the base-station. Despite several security enhancements in 5G networks, the rogues base-station, or equivalently MitM attacker, is still considered as serious threat for the network.

An adversary can exploit the coexistence environment to form a position like MitM (pseudo MitM) with similar attack capabilities. By taking advantage of the hidden node issue, an attacker can generate interference for the base-station during uplink transmissions while it also receives the messages of users. A similar attack can target selected user devices in download transmissions. Then, the attacker can replay the messages selectively to the base-station/device. This is slightly different from the classical MitM where the attacker intercepts the entire communication between the users and base-station. 

An adversary can also employ a similar mechanism, discussed above, to facilitate the deployment of a rogue base-station. The attacker blocks the communication channels at the base-station, by generating interference, while communicating with the users. During this time, the attacker can convince the users to connect to the rogue base-station. In none of these scenarios the base-station cannot distinguish between interference from a legitimate network sharing the spectrum or malicious interference of the attackers

\subsubsection{Physical Layer Security}

Existing security specifications do not include standardized protocols for physical layer security. However, there is an ongoing research and interest in defining authentication protocols that exploit RF features as a provision for physical layer security. Prominent examples of these techniques in the new radio (NR) physical layer of 5G and WiFi 6 are RF fingerprinting and beamforming for unique identification of devices. While these solutions seem promising, there are still challenges to be addressed including uncertainty and variability over device, time, and RF propagation channels. 

The physical layer security mechanisms are still targeting standalone networks and coexistence environments challenge their effectiveness. For instance, these techniques can provide additional identification information for authenticating devices within a network. However, unknown devices from outside the network domain must still be considered as legal in coexistence environments. Like other attacks, this can reduce the effectiveness of such security mechanisms.

\begin{figure}[t!]
	\centering
		\includegraphics[width=0.49\textwidth]{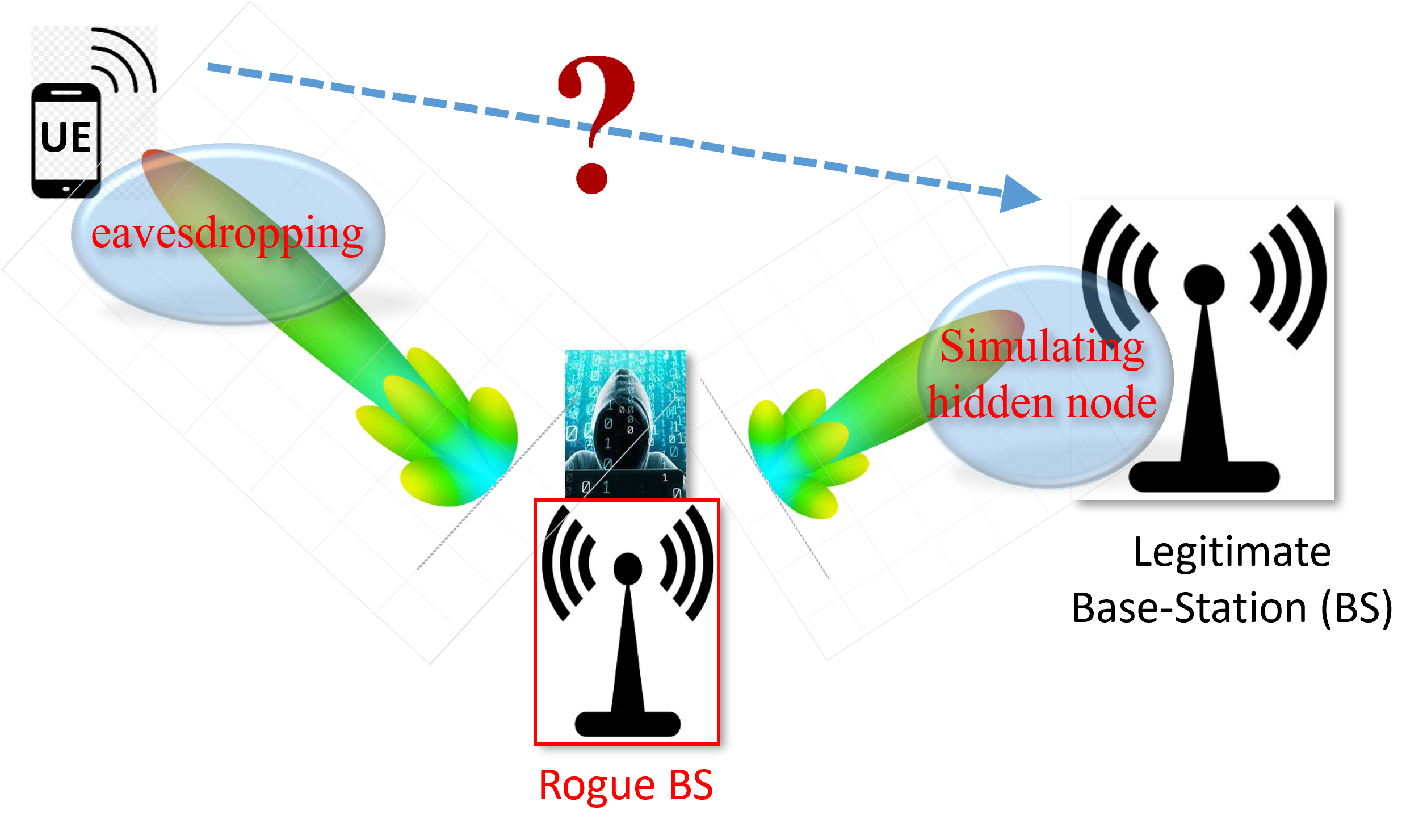}
	\caption{Exploiting beamforming and coexistence network environment in deploying effective MitM position.}
	\label{fig:beamforming}
\end{figure}

Ironically, the physical layer mechanisms proposed as security provisions can also help adversaries in deploying more effective attacks. In combined attacks exploiting the coexistence environment and the NR features, protection against such attacks becomes more challenging. In the example shown in Fig. \ref{fig:beamforming}, an attacker can effectively use beamforming to simulate a hidden-node interference for a legitimate base-station without affecting the user communication. Meanwhile, the attacker also eavesdrops on the communication from the victim device by forming another beam in the appropriate direction. In this process, the base-station cannot receive the messages from the user as it is experiencing interference. It also cannot distinguish the interference as malicious or legitimate due to coexistence conditions. Further, the user is not aware of the presence of the attacker. 

To further complicate the above attack, adversaries can employ RF fingerprinting to uniquely identify their target user devices. Hence, an attacker does not even need the device identifiers, such as 5G-GUTIs, to track the activity of the victim and eavesdrop on its communication. This example demonstrates how NR features, such as beamforming and OFDM modulations (exploited in RF fingerprinting) in a coexistence environment can compromise the security of individual networks. Hence, the importance of a secure mechanism for spectrum sharing is undeniable.

\section{Mitigation Plans and Future Research Direction}

\begin{figure*}[t!]
	\centering
		\includegraphics[width=0.9\textwidth]{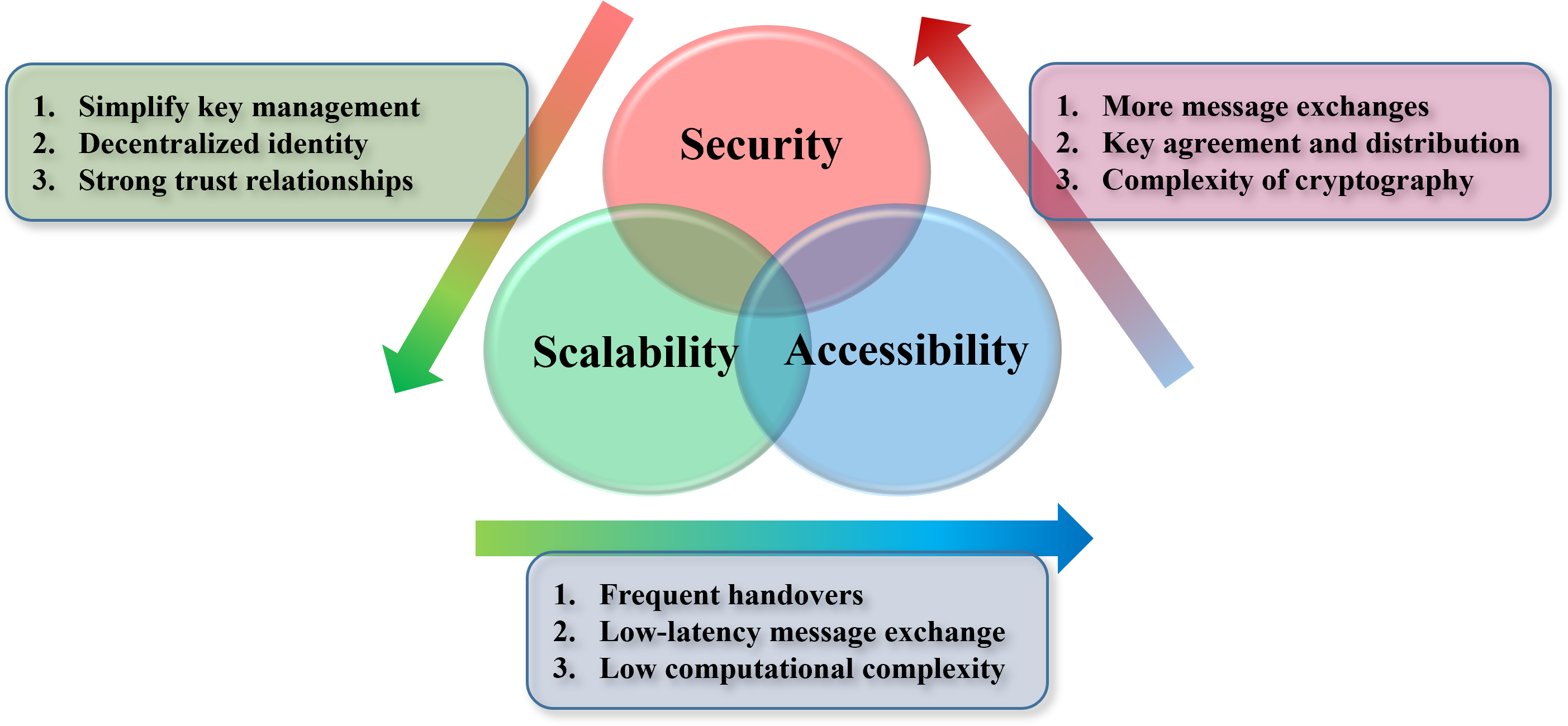}
	\caption{Trilemma often encountered in consolidating security while providing desired QoE in dynamic wireless networks.}
	\label{fig:trilemma}
\end{figure*}

Addressing security challenges in highly dynamic environments of next-generation wireless networks under QoE constraints is a multi-faceted optimization problem. A trilemma often encountered in this problem is shown in Fig. \ref{fig:trilemma}. A similar trilemma is also discussed in the context of blockchain security (by replacing accessibility with decentralization) \cite{grabe2020not, ghiro2021blockchain}. Seamless connectivity (accessibility), through a multi-RAT technology, is a main promise of 5G/6G networks. Scalability is an inevitable property of wireless networks as they are expected to support a massive volume of mobile devices in the era of IoT. Providing security while maintaining privacy is expected to incur minimal communication and computational overhead on network and devices. Improving any two of these properties could require a compromise in the third one, especially in the context of current static security frameworks. 

Current solutions for existing security issues often compromise one of the properties in the trilemma of Fig. \ref{fig:trilemma}. They might require more message exchanges, or more complex cryptography algorithms, for consolidated security in authentication and identification which increases latency and degrades the accessibility property. Part of the solutions requires pre-shared secret keys which is a challenge for billions of devices expected to operate in the network (scalability challenge). Next-generation cyber-security models based on zero trust architectures (ZTA), as discussed in \cite{ramezanpour2021intelligent}, may provide a mechanism to reach the optimal point (sweet spot) in this trilemma.

The more critical issue with existing security solutions is the assumption on the ownership of all network resources and negligence of coexistence characteristics, especially in the medium access control (MAC). The latter is a serious issue as any solution shall tolerate unknown accesses from co-existing networks. Furthermore, due to strict privacy constraints, identity of networks and the corresponding confidential information and usage requirements, cannot be shared among network operators. Hence, solutions based on cooperative medium access are challenging in practice. Hence, a secure coexistence framework should deal with unknown accesses while protecting the security of individual network users. In the following, we first review some existing proposals for addressing the security issues of standalone networks, discussed in the previous sections. Next, we will discuss the requirements of acceptable solutions to address security challenges in coexistence environments.

\subsection{Physical Layer Security Solutions}
Encrypted physical layer, as discussed in Section \ref{sec:physecurity}, can provide security with cryptographic proofs in protecting analog signals from an intercept. However, such solutions encounter scalability issues, in key agreement and distribution, as wireless networks are expected to support a substantially growing number of devices. Key exchange protocols such as Diffie-Hellman (D-H) can help in generating secret keys, however, the need for public-key infrastructure and associated certificate authorities is still an issue. The situation is exacerbated considering a large number of private 5G nano base-stations and WiFi access points in next-generation networks. Further, the D-H protocol, and the associated public-key cryptography, incurs large communication and computational overhead, especially on resource-constraint devices.

The secrecy capacity of communication channels, especially with beamforming in the NR physical layer of 5G networks, can be exploited to realize a key agreement protocol. This approach has been used in \cite{im2015secret} for a time-division duplex system with antenna arrays in the base-station and in the presence of pilot contamination. The base-station transmits random sequences to legitimate users while eavesdroppers attempt in forging the training signals of the users to cause deviation of the beam direction. The base-station then uses the eavesdropper transmissions to estimate information leakage and adjusts the length of secret keys accordingly. A similar approach is employed in \cite{im2015secret} with a two-way training (both uplink and downlink) which demonstrates improvement in estimating eavesdropper channel, hence, the efficiency of the key agreement protocol. Random pilot transmission for detecting active eavesdropper channels has also been used in other works such as \cite{tian2017random, tugnait2018pilot}.

Introducing artificial noise (AN) and matched filter precoding, in multi-input multi-output (MIMO) systems, is a common defense against pilot contamination \cite{zhu2014secure, wu2016secure}. Further, the null-space technique introduced in \cite{wu2016secure} help alleviate the exploitation of an eavesdropper by using the correlation diversity of user antennas. It is shown that under certain orthogonality conditions, this technique prevents an eavesdropper from reducing the secrecy rate. These techniques often require a perfect knowledge of the channel state information (CSI). A semi-blind technique is employed in \cite{hu2018secure} to estimate the legitimate users' signal which does not require CSI. The channel is estimated using data signals. A similar data-aided technique is used in \cite{wu2019data} to estimate the uplink channel during the training phase in massive MIMO systems.

Physical layer authentication (PLA) consolidates access security by incorporating unique characteristics of analog communication channels and/or devices in authentication protocols. A survey on various PLA techniques, based on CSI, frequency and identity watermarks, is provided in \cite{bai2020physical}. A combination of pre-shared secret keys and CSI for implementing a challenge-response protocol is used in \cite{choi2018coding}.  Fingerprint embedding is also a common technique in PLA \cite{perazzone2019physical}. Using artificial noise (AN) with imperfect knowledge of CSI is studied in \cite{perazzone2021artificial} for hiding a Hash-based message authentication code (HMAC). Employing angle-of-arrival (AoA) in realizing a PLA technique is investigated in \cite{xiong2013securearray}. Theoretical bounds on the performance of base-stations using AoA information of legitimate users, for estimating eavesdropper channels, have also been studied in \cite{darsena2020design}. 

\subsection{Access Security Solutions}
To provide perfect forward secrecy in layer 2 authentication mechanism, \cite{arkko2015usim} proposed integrating a D-H key exchange protocol into the 5G-AKA. This scheme also protects session keys from a passive eavesdropper with a knowledge of the long-term secret keys. This technique prevents revelation of the challenge nonce in the 5G-AKA protocol, hence the session keys, in passive eavesdropping attacks. The session key is generated from the long-term secret key and the challenge nonce. Employing a key agreement exchange at the beginning of authentication is a similar approach as used in the WiFi Protected Access (WPA3) mechanism which uses Dragonfly handshake rather than D-H \cite{vanhoef2020dragonblood}.

The use of D-H key exchange is also proposed in \cite{liu2018toward} for protecting against identity disclosure and replay attacks. The session key generated in the D-H handshake is used to encrypt the identifiers which guarantees identity confidentiality. Further, message exchanges with the 5G core network are accompanied by message authentication code (MAC) which prevents forging the messages and protects against replay attacks. Encryption of messages SYNC\_FAIL and MAC\_FAIL can also prevent traceability attacks which were discussed in Section \ref{sec:traceability}.

A common solution in the literature to address the challenges of public-key mechanisms, especially in establishing trust, is using blockchain as proposed in \cite{yang2017blockchain}. Combination of blockchain technology with public-key infrastructure (PKI) certificates has become a popular solution for implementing a lightweight and trusted platform for access security especially in device-to-device and vehicular communications \cite{boubakri2017access}. A blockchain structure for recording the certificates associated with access privileges of users in vehicular ad hoc networks (VANET) is introduced in \cite{lu2019blockchain} which also guarantees the identity privacy of users. A similar approach has also been developed based on Ethereum blockchain in \cite{lin2020bcppa}.

Cross-layer authentication techniques have also attracted attention as a means of protecting identity and location confidentiality, preventing message forgery, eavesdropping, and rogue base-stations. These techniques integrate PLA (as discussed in the previous section) with the layer 2 authentication mechanism. A review of cross-layer authentication mechanisms using PLA is presented in \cite{wang2016physical}. In most of these techniques, a physical layer characteristic is used as a fingerprinting parameter in the layer 2 authentication mechanism for initial identification and randomness generation \cite{tang2019light}. This approach is used in \cite{ma2019cross} with CSI and in \cite{moreira2018cross} with receive signal strength (RSS) as the fingerprinting features. 

To address the dynamic nature of 5G networks, \cite{zhang2020fast} introduces a coupled cross-layer mechanism between the PLA and upper-layer authentication mechanisms. This work employs a PLA with multiple fingerprinting features of the physical layer for higher reliability and stability. Further, the upper-layer authentication mechanism is also used to update the model parameters of the PLA for adaptation to the environment with low computational complexity. This is in contrast to the decoupled approach in \cite{pan2017physical} which employs PLA after a successful layer 2 authentication.

\subsection{Research Directions for Coexistence Security}
Ongoing research on security solutions, specifically for physical layer and access control, focuses on standalone networks. However, security in coexistence network environments requires revisiting secure access mechanisms that can tolerate out-of-network accesses. An important requirement of coexistence access security is preserving the privacy of individual networks. Employing a unified authentication and access control is also challenging as it either requires a trusted third party infrastructure or can introduce new security and privacy breaches.

A unique characteristic of coexistence network environment is the independent medium access control (MAC) protocols. Although the MAC protocols in 3GPP specifications for 5G networks and WiFi 6/6E are converging to similar algorithms, standalone networks have no means of detecting an activity in the spectrum as malicious from an attacker or legitimate from co-existing network users. The different and decoupled MAC mechanisms lead to hidden node conditions which cannot be prevented by intra-network mechanisms. The hidden node conditions can cause serious security exploits as discussed in Section \ref{sec:coexistence}. Hence, a coexistence security solution shall provide a mechanism for detection and avoidance of hidden node conditions.

Blockchain has been employed as a database spectrum sharing mechanism as a solution to the lack of unified MAC layers. This is in contrast to sensing-based spectrum sharing in which users make decisions on transmission based on their own, either individual or cooperative, measurements of the spectrum. The \textit{immutability} and \textit{transparency} properties of a blockchain can prevent non-legit users from accessing the spectrum. Further, the \textit{anonymity} property of a blockchain preserve the privacy of legitimate users accessing the spectrum. A review of the main properties of blockchains, and its applications in implementing distributed databases is provided in \cite{ghiro2021blockchain}. Based on these properties, \cite{weiss2019application} has introduced a fine-grained spectrum sharing, based on blockchain as a distributed database, for licensed spectrum access (LSA). Similarly, the application of blockchain for spectrum sharing in 5G machine-to-machine (M2M) communications is introduced in \cite{zhou2020blockchain}.

A major limitation of blockchain, and database spectrum sharing in general, is the requirement for a network access to the database. This is challenging in ad hoc wireless networks where devices require access to the spectrum for network connection. However, in wireless networks with centralized control, such as 5G and WiFi, the base-station might authorize access to the spectrum and schedule channels for mobile devices. Even in these networks, an adversary in a hidden node condition with respect to the base-station, can still cause interference for the mobile devices. The adversary can emulate the hidden node, e.g., by beamforming targeting the devices. Part of these vulnerabilities was discussed Sections \ref{sec:hijack} through \ref{sec:mitm}.

Higher levels of interference is also an expected feature of coexistence network environments. This situation promotes similar security challenges as non-orthogonal multiple access (NOMA) systems \cite{liu2017enhancing, ding2017spectral}. An attacker can exploit this environment to increase the interference levels without being identified as malicious. This can result in significantly degraded secrecy capacity in channel coding. Further, such covert attackers can compromise the security of key exchange protocols exploiting the secrecy capacity and PLA techniques using CSI and RF channel fingerprinting as discussed in the previous section. Dealing with malicious interference and distinguishing between attackers and legitimate users is an open problem with significant implications for coexistence security.

The security of beamforming physical layers and massive MIMO systems can also be challenged significantly in coexistence environments. As discussed in Section \ref{sec:beamforming}, pilot contamination attacks can cause the beams of antenna arrays to deviate from intended directions. coexistence environments provide attackers with unprecedented opportunities in deploying pilot contamination attacks. Existing defense mechanisms based on detecting and estimating eavesdropper channels rely on the assumption that all transmissions except for the eavesdropper are legitimate and follow the known protocols. However, in the coexistence environment, the interference from legitimate users of co-existing networks might cause the same or even stronger interference than an eavesdropper.

\section{Conclusion}
coexistence network environments introduce unique security challenges that have not been addressed within intra-network security frameworks and protocols. The primary assumption of standalone networks that all accesses shall be authenticated and authorized by the intra-network mechanisms is not valid anymore in coexistence environments. Unique characteristics of such network environments provide attackers with unprecedented exploits to degrade network capacity substantially and facilitate deployment of rogue base-stations. 

We reviewed key exploits that can result in serious security vulnerabilities with co-existing networks. A major challenge of coexistence is independent medium access control (MAC) in individual networks which results in hidden node conditions. An attacker can exploit this condition to access the spectrum, either to simply occupy a large portion of the spectrum or to increase interference levels, without being distinguished from a legitimate user. The first effect of such a simple attacker is a substantial drop in the network capacity. Further, deployment of rogue-base stations, man-in-the-middle attacks, and replay messages is facilitated in this environment. This study shows the necessity of revisiting existing security solutions by taking the specific characteristics of coexistence environments into consideration for the next generation of networks. 

 \bibliographystyle{elsarticle-num} 
 \bibliography{MRLab}





\end{document}